\documentclass[10pt,amsmath,amssymb,aps,pra,twocolumn]{revtex4-2}

\usepackage{graphicx}
\usepackage{dcolumn}
\newcolumntype{d}[1]{D{.}{.}{#1}}

\usepackage{upgreek}

\usepackage{physics}

\newcommand{\boldgreek}[1]{{\mbox{\boldmath$ {#1} $}}}

\usepackage{xcolor}

\newcommand{\Yb}[1]{${}^{#1}\mathrm{Yb}^+$}
\newcommand{\Bc}{B_\mathrm{crit}}

\begin{document}

\title{Suppression of differential light shifts in ground and metastable trapped-ion qubits}

\author{Drew Parks}\thanks{These authors contributed equally to this work.}
\affiliation{University of California Los Angeles, Los Angeles CA USA 90095}

\author{Thomas Dellaert}\thanks{These authors contributed equally to this work}\thanks{Presently at IonQ.}
\affiliation{University of California Los Angeles, Los Angeles CA USA 90095}

\author{Patrick McMillin}\thanks{Presently at Quantinuum.}
\affiliation{University of California Los Angeles, Los Angeles CA USA 90095}

\author{Conrad Roman}\thanks{Presently at Quantinuum.}
\affiliation{University of California Los Angeles, Los Angeles CA USA 90095}

\author{Andrei Derevianko}
\affiliation{Department of Physics, University of Nevada, Reno, Nevada 89557, USA}

\author{Wesley C. Campbell}
\affiliation{University of California Los Angeles, Los Angeles CA USA 90095}

\date{\today}

\begin{abstract}
In the presence of a magnetic field, hyperfine clock qubits can acquire a vector differential light shift that can be tuned via polarization to suppress the total differential light shift of high-power, off-resonant laser light. We experimentally measure this ``magic'' polarization condition, suppressing differential light shifts in both the ${}^2\mathrm{S}_{1/2}$ ground and  ${}^2\mathrm{F}_{7/2}^o$ metastable clock qubits of \Yb{171}. We present calculations of the minimum bias magnetic fields required to suppress differential light shifts in the ground state clock qubits of commonly trapped ion species, finding that they are below the strengths of fields already typically present in experiments. We further present methods for metastable clock-qubit control in \Yb{171}, demonstrating a state preparation and measurement infidelity of $2.9^{+3.0}_{-1.5}\times10^{-4}$ ($-35 \pm 4 \, \mathrm{dB}$).
\end{abstract}

\maketitle

In trapped-ion qubit systems, high-power off-resonant laser light is an often-used method for performing single- and two-qubit gates \cite{PhysRevLett.117.060504, PhysRevLett.117.060505}. However, this off-resonant laser light can induce undesirable differential light shifts between the qubit states.
These light shifts can map gate-laser intensity noise to the qubit frequency, resulting in dephasing and reducing the effective coherence time. 
Further, in operational schemes that rely on the isolation of qubit subspaces such as the $omg$ architecture \cite{omgBluePrint}, these differential light shifts may cause detrimental crosstalk errors \cite{BariumMagic}, in which qubits not participating in the interaction experience phase shifts from the perturbing light.

Magnetic-field-induced vector differential light shifts \cite{lundbladExperimentalObservationMagicwavelength2010} have been shown to give rise to a ``magic''  polarization condition in neutral atoms that suppresses this source of decoherence \cite{PhysRevLett.104.073604,NuetralMagic}. Recently, this magic condition was demonstrated in the ground state qubits of trapped ionic species ${}^{133}\mathrm{Ba}^+$ \cite{BariumMagic} and ${}^{171}\mathrm{Yb}^+$ \cite{YtterbiumMagic} as well. In this paper, we explore extensions of magic polarization to ground state qubits of various commonly trapped ions, and present experimental observations of the magic polarization effect in both the ground and metastable $m_F=0$ clock qubits of \Yb{171} in magnetic field environments typical of trapped-ion experiments. 

\section*{\centering Theory}
\label{Sec:Theory}

We consider the  clock qubit encoded in the
$nS_{1/2}$ ground state hyperfine manifolds characteristic of many commonly used trapped-ion
species with non-zero nuclear spin $I$. This manifold consists of two hyperfine components. We focus on the two B-field-insensitive qubit clock states
$\ket{F} \equiv |F,m_F=0\rangle$ and $\ket{F'} \equiv  |F',m'_F=0\rangle$, where $F'$ is the upper and $F$ is the lower energy state. These are eigenstates for unperturbed Hamiltonian $\hat{H}_0$. For the positive hyperfine structure constant $A_{nS_{1/2}}$, $F'=I+1/2$ and $F=I-1/2$ and the assignment is reversed otherwise. In the absence of applied fields, the qubit angular frequency $\omega_q = E_{F'}- E_{F} > 0$ ($\hbar \equiv 1$). 

Below we review the derivation of the magic conditions, where both qubit levels experience identical perturbations by the laser field. This section is based on the earlier work on magic trapping of neutral atoms~\cite{RosGheDzu09,BelDerDzu09Clock,ChiNelOlm10,Der10Bmagic}, which was recently extended to laser-noise insensitive operation of $^{133}$Ba$^+$ qubit~\cite{BariumMagic}. We also justify certain approximations made in the earlier work.

Interaction of an ion with a weak laser field of frequency $\omega_L$, amplitude $\mathcal{E}_L$ and polarization $\boldgreek{\hat{\varepsilon}}$ can be parameterized by an optical potential operator
\begin{equation}
    \hat{U} = - \hat{\alpha}(\omega_L, \boldgreek{\hat{\varepsilon}}) \frac{\mathcal{E}^2_L}{4} + \mathcal{O}(\mathcal{E}^4_L) \,, 
    \label{Eq:OptPot}
\end{equation}
where $\hat{\alpha}(\omega_L, \boldgreek{\hat{\varepsilon}})$ is the operator of dynamic polarizability that can be expanded into irreducible tensor operators of 
rank $K=0,1,2$,
\begin{equation}
 \hat{\alpha}(\omega_L, \boldgreek{\hat{\varepsilon}}) = \hat{\alpha}^{(0)}(\omega_L) + \mathcal{A}  \hat{\alpha}^{(1)}(\omega_L)  + \hat{\alpha}^{(2)}(\omega_L)\,,
\end{equation}
where $\mathcal{A}\equiv-i\left(\boldgreek{\hat{\varepsilon}}^* \times \boldgreek{\hat{\varepsilon}}\right) \cdot \mathbf{\hat{z}}$ is the photon's helicity projection on the quantization axis $\mathbf{\hat{z}}$, and AC polarizabilities $\hat{\alpha}^{(K)}$ are proportional to the conventional scalar, vector and tensor polarizabilities~\cite{RosGheDzu09}.

Importantly, since the laser field breaks the rotational invariance of atomic Hamiltonian $\hat{H}_0$, 
$\hat{\alpha}(\omega_L, \boldgreek{\hat{\varepsilon}})$ can have both diagonal and off-diagonal matrix elements in the $\{ \ket{F}, \ket{F'}\}$ basis. In an externally applied magnetic field, energies of qubit states can be obtained by diagonalizing the $\hat{H}_0 + \hat{U} +\hat{H}^Z$ matrix, where $\hat{H}^Z$ is the Zeeman Hamiltonian. We will take the quantization axis along the laser propagation direction $\mathbf{\hat{k}}$. In our treatment (and in the experiment), the B-field is also directed along $\mathbf{\hat{k}}$. Then $\hat{H}^Z = - \hat{\mu}_z B$, where $\hat{\mu}_z$ 
is the magnetic dipole operator's $z$ component. While the diagonal matrix elements of $\hat{H}^Z$ vanish for our qubit states (there is no linear Zeeman shift for the $M_F=0$ states), it admixes the qubit basis states, leading to quadratic Zeeman shifts.

At zero magnetic field, the off-diagonal matrix elements of the optical potential lead to $\mathcal{O}(\mathcal{E}^4_L)$ corrections, which will be ignored for consistency with the omitted hyper-polarizability contributions in Eq.~\eqref{Eq:OptPot}. If the laser field is off, eigen-energies repel each other exhibiting quadratic Zeeman shift.
When both fields are on, the correction to the energy splitting can be estimated in the second-order perturbation theory
\begin{equation}
\Delta\omega_\mathrm{DLS} \approx U_{FF}-U_{F'F'}
+
\frac{4}{\omega_q}
\mathrm{Re}\!\left(H^Z_{FF'}U_{F'F}\right)\,, \label{Eq:DLS}
\end{equation}
where we omitted the quadratic Zeeman shift. This is the differential light shift (DLS) of interest. The DLS expression~\eqref{Eq:DLS} contains B-field-free DLS and a Zeeman-optical interference term.

We would like to find the conditions when the DLS vanishes, thereby eliminating the associated laser-induced decoherence. The DLS can be recast in terms of differential polarizability $\Delta\alpha(\omega_L, \boldgreek{\hat{\varepsilon}})$, so that
$\Delta\omega_\mathrm{DLS} = - \Delta\alpha(\omega_L, \boldgreek{\hat{\varepsilon}}) \, \mathcal{E}^2_L/4$. Here 
\begin{equation}
\Delta\alpha(\omega_L, \boldgreek{\hat{\varepsilon}})  \equiv \alpha_{F'F'}-\alpha_{FF}
- \frac{4 B \mu_\mathrm{B}}{\omega_q}
 \alpha_{F'F} \,, \label{Eq:DeltaAlpha}
\end{equation}
where we used $\mel{S_{1/2},F,M_F=0}{\hat{\mu}_z}{S_{1/2},F',M_{F'}=0} = \mu_\mathrm{B}$, with $\mu_\mathrm{B}$ being the Bohr magneton. We suppressed the $\omega_L$ and $\boldgreek{\hat{\varepsilon}}$ dependencies of polarizabilities on  the r.h.s.\ of this equation for brevity.

Since both qubit states are attached to the same electronic level, their polarizabilities are nearly identical, so that $\alpha_{F'F'} \approx \alpha_{FF}$. The small but important differences arise due to hyperfine interaction (HFI) admixtures to electronic wave functions and energies. The detailed formalism for computing  HFI-mediated corrections $\hat{\beta}^{(K)}$ to the AC polarizabilities can be found in Ref.~\cite{RosGheDzu09} which generalizes the earlier static polarizability derivation~\cite{BelSafDer06}. Such HFI-mediated corrections are derived in third order of Floquet perturbation theory and involve matrix elements of HFI and two dipole operators along with energy denominators. Three types of diagrams appear in the calculations depending on the position of the HFI operator in the chain of three operators: top, center, bottom, see Fig.~1 of Ref.~\cite{RosGheDzu09}. The third-order diagrams also include normalization corrections. We do not reproduce the lengthy formulae from Ref.~\cite{RosGheDzu09} here, but these are the expressions that we use in our calculations. Note that simply including hyperfine splittings in the energy denominators while neglecting hyperfine corrections to the electronic wave functions can lead to qualitatively erroneous results~\cite{RosGheDzu09}, because the two effects enter at the same order.

In general, both the scalar ($K=0$) and tensor ($K=2$) HFI-mediated polarizabilites $\hat{\beta}^{(K)}$ contribute to the $\alpha_{F'F'}-\alpha_{FF}$ term in Eq.~\eqref{Eq:DeltaAlpha}. For ground-state alkali-metal atoms, scalar 
contribution $\hat{\beta}^{(0)}$ strongly dominates~\cite{BelDerDzu09Clock}. The same inequality  $\beta^{(0)} \gg \beta^{(2)}$ also holds for the alkali-like $S_{1/2}$ 
ions considered here. A detailed analysis reveals that $\beta^{(2)}$ is dominated by the center diagram which requires HFI matrix elements evaluated between the $P_J$ intermediate states, while $\beta^{(0)}$ is largely determined by the normalization (residual) diagram involving a much larger expectation value of HFI in the $nS_{1/2}$ ground state.
The normalization diagram vanishes identically for the $\beta^{(2)}$ tensor polarizability. Thus $\beta^{(0)} \gg \beta^{(2)}$ and we neglect tensor contribution, so that $\alpha_{F'F'}-\alpha_{FF} \approx \beta_{F'F'}^{(0)} - \beta_{FF}^{(0)}$ in Eq.~\eqref{Eq:DeltaAlpha}. 
To quantify systematic errors associated with this approximation, we carried out numerical tests for the Yb$^+$ ion  which has a softer core compared to other ions of interest and find that  neglecting 
tensor polarizability introduces the largest few-percent fractional errors near the $6S_{1/2}-6P_J$ resonances. These two scalar polarizabilities are 
strictly proportional to each other precluding B-field-free magic wavelengths for the $S_{1/2}$ hyperfine manifolds~\cite{BelDerDzu09Clock}. This is where the Zeeman-optical interference term in Eq.~\eqref{Eq:DLS} becomes important as it can cancel out the residual B-field-free DLS. 

The Zeeman-optical interference term can be expressed in terms of the traditional vector (axial) polarizability $\alpha_{nS_{1/2}}^{a}\left(  \omega_{L}\right)$, 
as  the off-diagonal matrix element of the scalar polarizability $\hat{\alpha}^{(0)}$ vanishes due to the angular selection rules and the tensor $\hat{\alpha}^{(2)}$ contribution for the $S_{1/2}$ states arises only through the suppressed HFI admixtures.

Combining the terms in Eq.~\eqref{Eq:DeltaAlpha}, we find that  DLS vanishes at the ``magic'' helicity projection
\begin{equation}
\mathcal{A}_{m}\left(  \omega_{L}\right)  \approx-\frac{2I+1}{2I}%
\frac{\beta_{FF}^{\left(  0\right)}\left(  \omega_{L}\right)
}{\alpha_{nS_{1/2}}^{a}\left(  \omega_{L}\right)  }~\frac{\hbar \omega_q}{\mu_\mathrm{B} B} \equiv -\frac{\Bc(\omega_L)}{B} \,,\label{Eq:MagicA}%
\end{equation}
where $\beta_{FF}^{\left(  0\right)}\left(  \omega_{L}\right)$ is the  HFI-mediated scalar polarizability of the lower qubit clock state. Here we defined the critical magnetic field
\(\Bc\) as the minimum field for which the DLS can
be canceled by an allowed, \(|\mathcal{A}|\leq 1\), polarization. Fields
larger than $\Bc$ allow cancellation with partially circular
polarization, while fields below $\Bc$  cannot provide a sufficiently large Zeeman-optical interference  contribution to compensate the scalar differential shift in Eq.~\eqref{Eq:DeltaAlpha}.

The ratio $\beta_{FF}^{\left(  0\right)}\left(\omega_{L}\right)  /\alpha_{nS_{1/2}}^{a}\left(  \omega_{L}\right)  $ is on
the order of the ratio of the hyperfine splitting $\omega_q$ to the fine-structure
splitting in the lowest energy $P$-state manifold. In addition, the vector polarizability $\propto \omega_L$, and 
we expect $\Bc(\omega_L)$ to scale as $1/\omega_L$ below the lowest-energy $nS_{1/2} - nP_{1/2,3/2}$ resonance.

If the B field and the direction of laser propagation are set at an angle $\theta$,
then $\mathcal{A}$ in Eq.~\eqref{Eq:MagicA} is proportional to $\cos \theta$, providing an 
additional experimental handle. Increasing the angle and reducing
the degree of circular polarization raises the critical B field.

The DLS cancellation is independent of the laser intensity:
intensity sets the magnitude of the uncompensated differential light
shift, while the ratio of scalar to vector polarizability determines the
polarization required for cancellation. At the magic polarization, the two
clock states experience equal light shifts from the perturbing laser,
so fluctuations in the laser intensity no longer map directly onto
fluctuations of the qubit state splitting.

\subsection*{Numerical evaluation of critical fields}

We carried out critical-field calculations for the $S_{1/2}$
ground state hyperfine clock manifolds of several commonly used trapped-ion
species: $^{171}\mathrm{Yb}^+$, $^{173}\mathrm{Yb}^+$,
$^{133}\mathrm{Ba}^+$, $^{135}\mathrm{Ba}^+$,
$^{137}\mathrm{Ba}^+$, $^{87}\mathrm{Sr}^+$,
$^{25}\mathrm{Mg}^+$, $^{43}\mathrm{Ca}^+$, and
$^{9}\mathrm{Be}^+$. All these isotopes have a non-zero nuclear spin $I$ tabulated in the supplemental materials along with the nuclear g-factors $g_I$.

The critical B field is expressed as (c.f.\ Eq.~\eqref{Eq:MagicA})
\begin{equation}
\Bc\left(  \omega_{L}\right)  = \frac{2I+1}{2I}%
\frac{\beta_{FF}^{\left(  0\right)}\left(  \omega_{L}\right)
}{\alpha_{nS_{1/2}}^{a}\left(  \omega_{L}\right)  }~\frac{\hbar \omega_q}{\mu_\mathrm{B}} \,.\label{Eq:MagicB}%
\end{equation}

Computations of vector polarizability $\alpha_{nS_{1/2}}^{a}\left(  \omega_{L}\right)$ of  the ground $nS_{1/2}$ state
involve a summation over intermediate opposite-parity  atomic states $\ket{i} = \ket{n_i P_{1/2,3/2}}$, electric-dipole matrix elements $\mel{nS_{1/2}}{\mathbf{D}}{n_i P_{1/2,3/2}}$, and energy  denominators  of both rotating
and counter-rotating form $E_g-E_i \pm \omega_L$ ($E_g$ is the ground state energy). The energy denominators lead to a resonant behavior of the polarizabilites. Determination 
of the HFI-induced scalar polarizability $\beta_{FF}^{\left(  0\right)}\left(  \omega_{L}\right)$ introduces an additional summation over intermediate states and requires magnetic-dipole HFI matrix elements, especially of the $S_{1/2}-S_{1/2}$ character for the top/bottom diagrams and $P_J-P_{J'}$ for the center diagrams. The required dipole matrix elements additionally include $P_{J}-D_{J'}$ channels.

We employ {\em ab initio} relativistic random-phase approximation and Brueckner orbital (RPA+BO) methods implemented as described in Ref.~\cite{tan2023precision}. Both RPA and BO are all-order (in residual electron-electron Coulomb interaction) many-body methods. To ensure the completeness of the intermediate state basis and to improve numerical accuracy in computing HFI, we employed the dual-kinetic-balance B-spline basis set~\cite{BelDer08.DKB}. These basis sets were generated in the $V^{N-1}$ frozen core Dirac-Hartree-Fock (DHF) potentials with a finite-size nucleus and further rotated by diagonalizing the DHF+BO Hamiltonian to generate numerically complete independent particle DHF+BO basis sets~\cite{tan2023precision}.

A typical accuracy of the relativistic RPA+BO method is sub-percent for light mono-valent ions such as Be$^+$ 
and a few percent for heavy ions. The most demanding is the case of Yb$^{+}$ with an easily 
polarizable core due to the outer $f$-shell. We present a comparison of the RPA+BO dipole matrix 
elements for Yb$^{+}$ with literature values in Table~\ref{Tab:Ybplus_E1_matrix_elements}. Notice that even more complete (than our RPA+BO) relativistic coupled cluster treatment substantially disagrees with experimental values for the $6s_{1/2} - 6p_{j}$ matrix elements. 

\begin{figure*}[p]
\includegraphics[width=.87\textwidth]{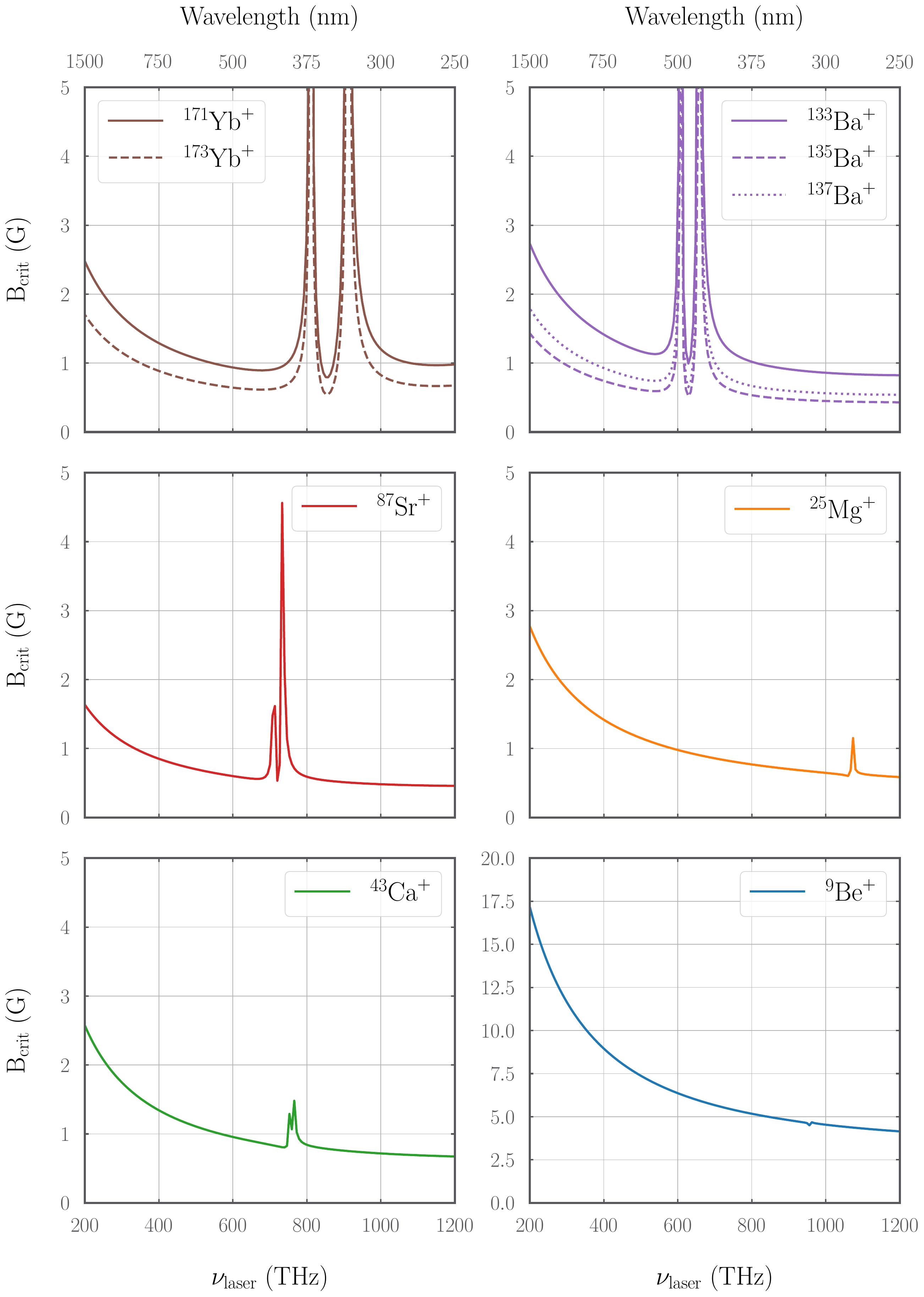}
\caption{Full-scale critical magnetic field calculations including the hyperfine interaction for various commonly trapped ions. Note that for most driving frequencies and ions, the critical field is within the range typical to ion trap operation ($O$(Gauss)).  The most-likely exception to this is ${}^9\mathrm{Be}^+$, for which the vertical scale has been expanded to show $B_\mathrm{crit}$ (bottom right).}
\label{BcritIons}
\end{figure*}

\begin{table}[t]
\caption{
Reduced electric-dipole matrix elements in Yb$^+$, in atomic units $ea_0$.
The RPA+BO values are the present {\em ab initio} relativistic values. MBPT denotes the
third-order relativistic many-body perturbation theory values of
Ref.~\cite{SafronovaSafronova2009YbII}, and RCC denotes the relativistic
coupled-cluster values of Ref.~\cite{SahooDas2011YbPNC}. Experimental entries
are direct or lifetime/branching-fraction inferred values from
Refs.~\cite{Shao2024YbDipole,PhysRevA.97.032511}. Signs of reduced matrix
elements are phase-convention dependent; magnitudes are listed.
}
\label{Tab:Ybplus_E1_matrix_elements}
\begin{ruledtabular}
\small
\begin{tabular}{
l
@{\hspace{0em}}D{.}{.}{1.3}
@{\hspace{0em}}D{.}{.}{1.3}
@{\hspace{0em}}D{.}{.}{1.2}
@{\hspace{0em}}D{.}{.}{1.3}
}
Transition &
\multicolumn{1}{l}{RPA+BO} &
\multicolumn{1}{l}{MBPT} &
\multicolumn{1}{l}{RCC} &
\multicolumn{1}{l}{Expt.} \\
\hline
$6s_{1/2}-6p_{1/2}$ & 2.667 & 2.683 & 2.72(1) & 2.470(3) \\
$6s_{1/2}-6p_{3/2}$ & 3.759 & 3.768 & 3.83(1) & 3.36(3) \\
$5d_{3/2}-6p_{1/2}$ & 3.047 & 2.971 & 3.06(2) & 2.998(2) \\
$5d_{3/2}-6p_{3/2}$ & 1.343 & 1.307 & 1.35(2) & 1.15(4) \\
$5d_{5/2}-6p_{3/2}$ & 4.216 & 4.120 & 4.23(3) & 3.95(10) \\
\end{tabular}
\end{ruledtabular}
\end{table}

To improve the accuracy of resonances we replaced our theoretical DHF+BO energies by the NIST recommended values~\cite{Kramida_NIST_ASD} for several lowest-lying energy levels of the $S_{1/2},P_J$ and $D_J$ symmetry. To further improve the accuracy, we used literature values for dipole matrix elements and hyperfine structure constants for low-lying states. The employed non-DHF+BO values are listed in supplementary material. Off-diagonal matrix elements 
of the $n's - n''s$ hyperfine interaction were computed using geometric mean of the $A_{n's}$ and $A_{n''s}$ hyperfine structure constants to substantially improve the accuracy~\cite{Derevianko1999b} of these HFI matrix elements.
The remaining matrix elements and energies were our RPA+BO values.

For each laser frequency, these scalar and axial polarizabilities were
inserted into Eq.~\eqref{Eq:MagicB}. The resulting values of $\Bc$ as a function of
the laser frequency are shown in Fig.~\ref{BcritIons}. 
For most wavelengths relevant to trapped-ion operation, the calculated
critical fields are in the range of a few gauss, comparable to bias fields
already used in typical ion-trap experiments to destabilize coherent dark states and resolve Zeeman structure. These 
$\Bc(\omega_L)$ curves reflect the ratio of a small
HFI-mediated scalar response to the ordinary electronic vector response.
Resonant features in Fig.~\ref{BcritIons} occur at strong
$S_{1/2}\rightarrow P_{J}$ electric-dipole transitions, where both
polarizabilities vary rapidly. $\Bc(\omega_L)$ scales roughly as $1/\omega_L$ below the lowest-energy $nS_{1/2} - nP_{1/2}$ resonance as expected.

The isotope dependence of the critical field follows from the HFI-mediated scalar polarizability and the clock frequency in Eq.~\eqref{Eq:MagicB}. 
We neglect hyperfine anomalies and small isotope shifts in the electronic matrix elements. Then at fixed laser frequency and within a given ion species, the electronic vector polarizability and electronic sums entering $\beta^{(0)}$ are approximately isotope independent. 

After angular reduction for the two $S_{1/2}$ hyperfine components~\cite{Der10Bmagic}, $\beta^{(0)}_{I+1/2}=-(I+1)\beta^{(0)}_{I-1/2}/I$ and $\beta^{(0)}_{I-1/2}\propto g_I I$, so the differential scalar response scales as $g_I(2I+1)$. The clock splitting also scales as $\omega_q \propto |g_I|(2I+1)$. Therefore
\begin{equation}
\label{B_crit_nuclear_scaling} \Bc(\omega_L)\propto g_I^2(2I+1)^2 \,.
\end{equation}
This implies that a measurement of critical field at fixed laser frequency in one isotope can be rescaled to another isotope of the same ion species.

\section*{Experiments}
\subsection*{Metastable Qubit Operations}

\begin{figure}
\includegraphics[width=.5\textwidth]{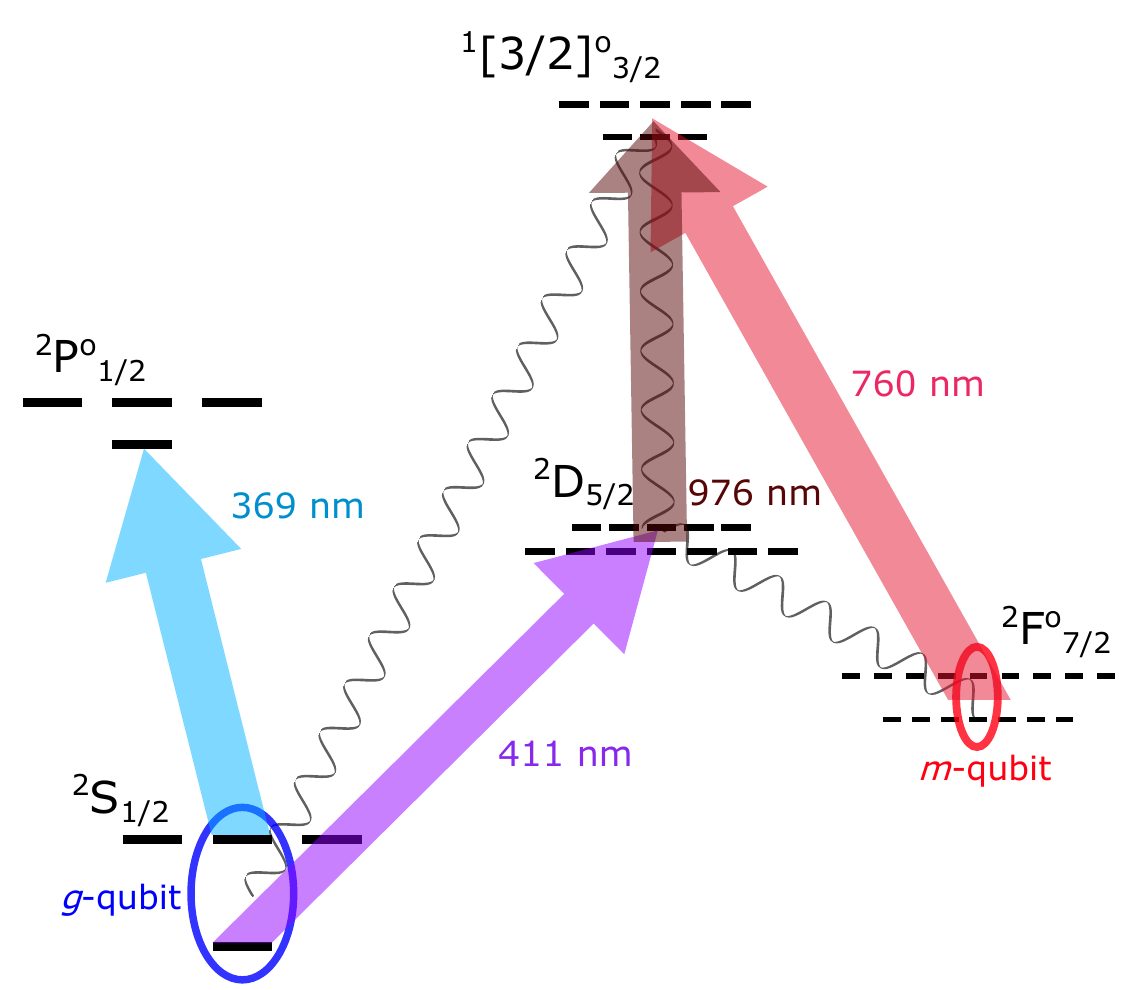}
\caption{The $369\,\mathrm{nm}$ laser controls Doppler cooling, optical pumping, and fluorescence state detection in the $^2\mathrm{S}_{1/2}$ ground qubit with two electro-optic modulator tones. The metastable qubit is prepared by driving an $E$2 (${}^2\mathrm{S}_{1/2}\;F = 0$ to ${}^2\mathrm{D}_{5/2}\; F=2$) transition at $411\,\mathrm{nm}$, which decays to $^2\mathrm{F}^o_{7/2}\; F=3$. Read out is performed with a $^2\mathrm{F}^o_{7/2}\;F=3$ to $^1[3/2]^o_{3/2}\; F=1$ $760 \,\mathrm{nm}$ transition, which decays to the $^2\mathrm{S}_{1/2}$ hyperfine manifold for fluorescence detection. The $976\,\mathrm{nm}$ laser acompanies the $760 \,\mathrm{nm}$ to repump population from ${}^2\mathrm{D}_{5/2}$ during readout operations. Qubit rotations for both the $g$- and $m$- type qubits are effected via microwaves applied by external horn antennas.}
\label{SPAM_grotrian}
\end{figure}

\begin{figure}
\begin{tabular}{cc}
\includegraphics[width=.45\textwidth]{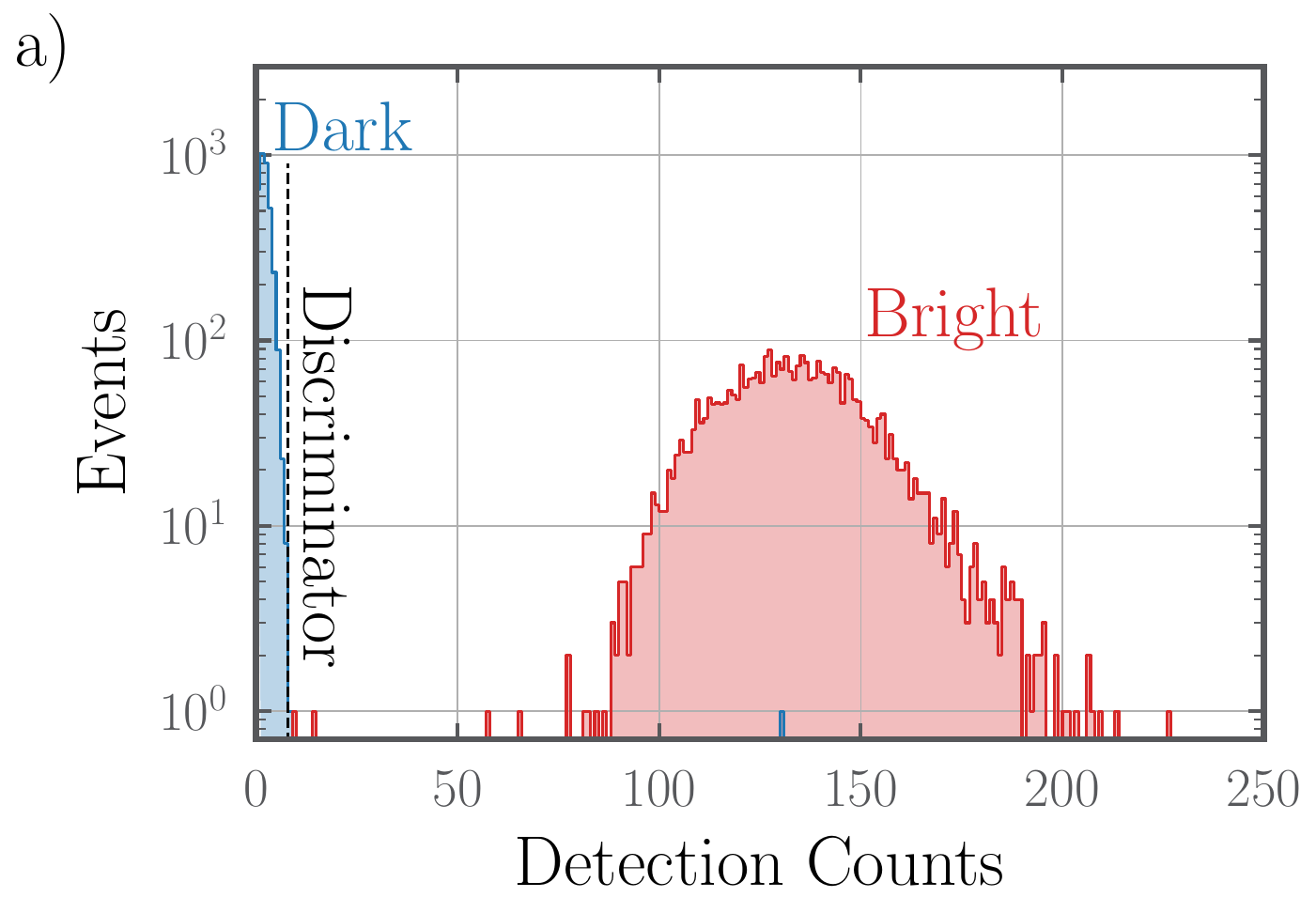}\\
\newline
\includegraphics[width=.45\textwidth]{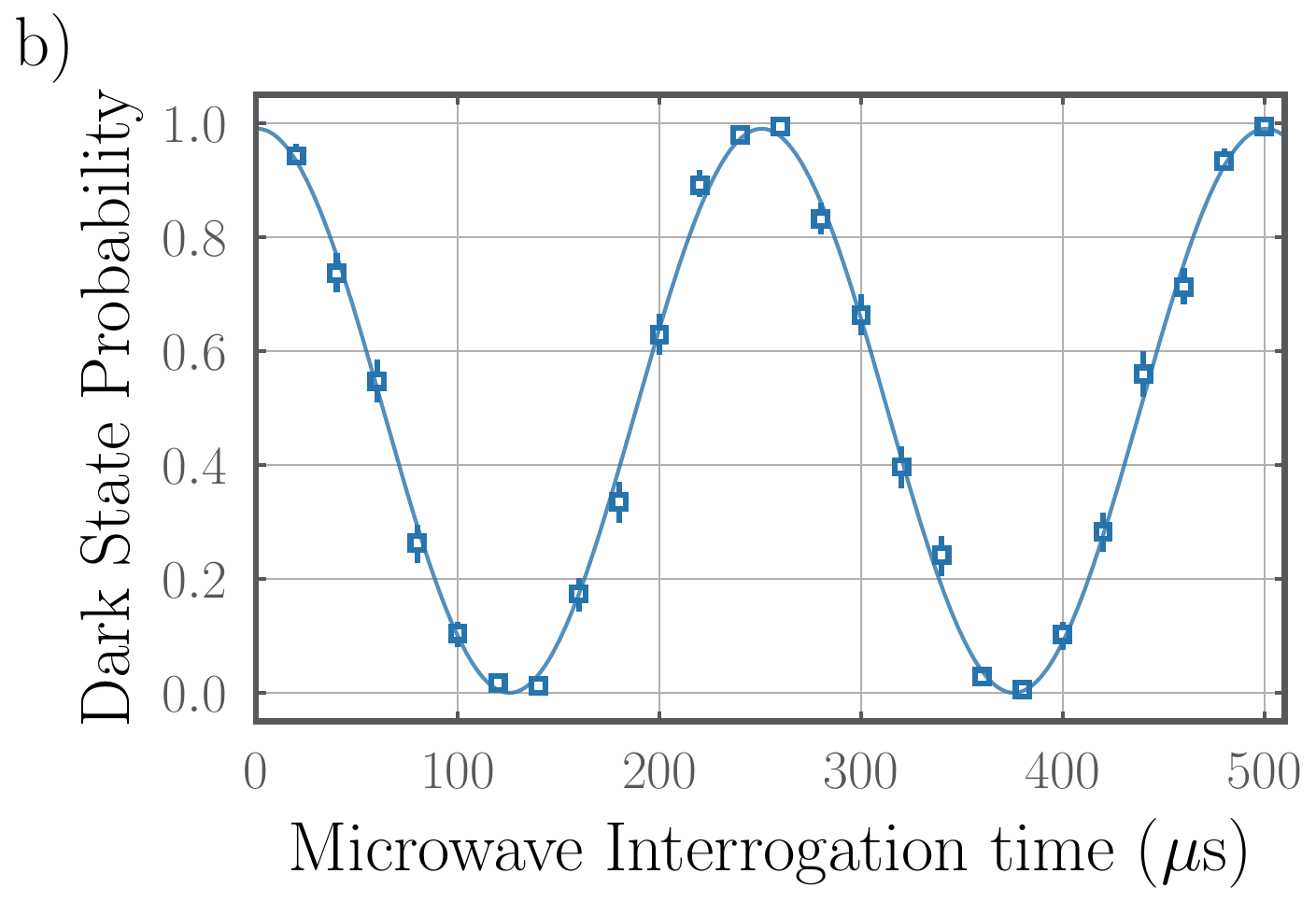}
\end{tabular}
\caption{a) Detected photon count histograms for prepared metastable bright and dark states during $3.5\,\mathrm{ms}$ of state detection time. b) Metastable-qubit Rabi flops show high contrast even without correcting for finite SPAM fidelity.}
\label{Mqubit}
\end{figure}

The metastable-state (``$m$-type'' \cite{omgBluePrint}) clock qubit is defined in the long-lived (3.16 years \cite{lange_lifetime_2021}) $^2\mathrm{F}^o_{7/2}$ hyperfine manifold between the $\ket{3,\,0}$ and $\ket{4,\,0}$ states (denoted $\ket{F,\,m_F}$), with zero-field qubit frequency splitting we measure to be $\omega_m=2\pi\times3.620\;\!527\;\!20(5)\,\mathrm{GHz}$. The metastable qubit is prepared via heralded optical pumping through a decay pathway from the $^2\mathrm{D}_{5/2}$ hyperfine manifold \cite{ransfordWeakDissipationHighfidelity2021}, with relevant states shown in figure \ref{SPAM_grotrian}. Starting in the ground-state Doppler cooling cycle, we optically pump into $^2\mathrm{S}_{1/2}\ \ket{0,\,0}$, then apply a $411 \,\mathrm{nm}$ laser for $100 \,\mathrm{ms}$ to drive the electric quadrupole ($E$2) transition to $^2\mathrm{D}_{5/2}\ \ket{2,\,\pm1}$.  This state preferentially decays to $^2\mathrm{F}^o_{7/2}\; F=3$, populating the $^2\mathrm{F}^o_{7/2}\ \ket{3,\,0}$ clock state with $\approx20\%$ probability. 

To prepare the $^2\mathrm{F}^o_{7/2}\ \ket{4, \,0}$ state, we drive a microwave $\pi$-pulse from $^2\mathrm{F}^o_{7/2}\ \ket{3,\,0}$, then drive any remaining population in $^2\mathrm{F}^o_{7/2}\; F=3$ to $^1[3/2]^o_{3/2}\; F=1$ with a $760 \,\mathrm{nm}$ laser applied for $50 \,\mathrm{ms}$. This excited state has a short lifetime ($29\,\mathrm{ns}$ \cite{Berends}), and preferentially decays to the $^2\mathrm{S}_{1/2}$ ground state \cite{Sugiyama_1999}. We then use ground-state Doppler cooling light to detect any population in the $^2\mathrm{S}_{1/2}$ manifold, which would indicate failed preparation of $^2\mathrm{F}^o_{7/2}\ \ket{4, \,0}$, in which case the procedure is repeated until successful state preparation is heralded by the lack of laser-induced fluorescence from the Doppler cooling light.

\begin{figure*}
\begin{tabular}{cc}
\centering
\includegraphics[width=.45\textwidth]{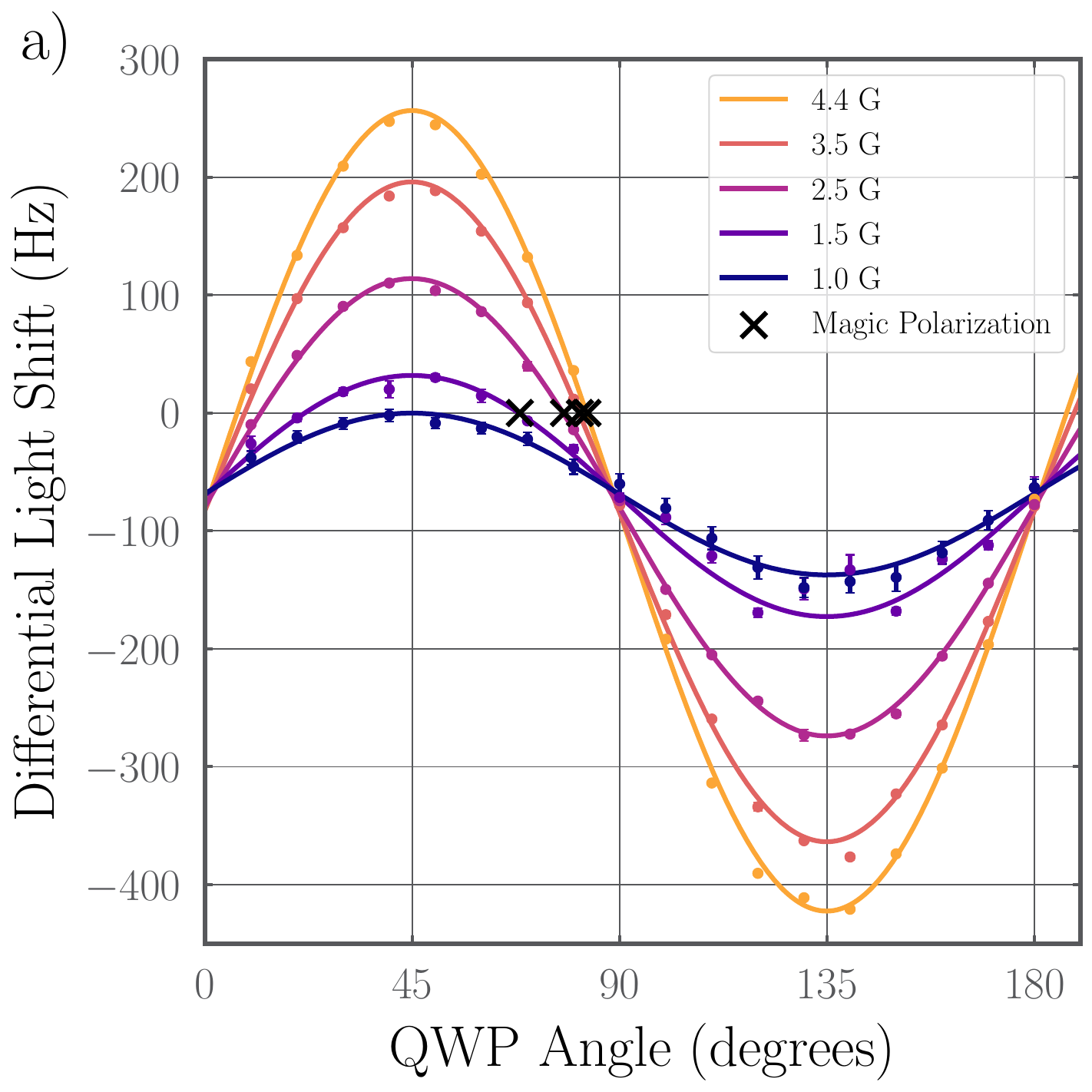} &\includegraphics[width=.45\textwidth]
{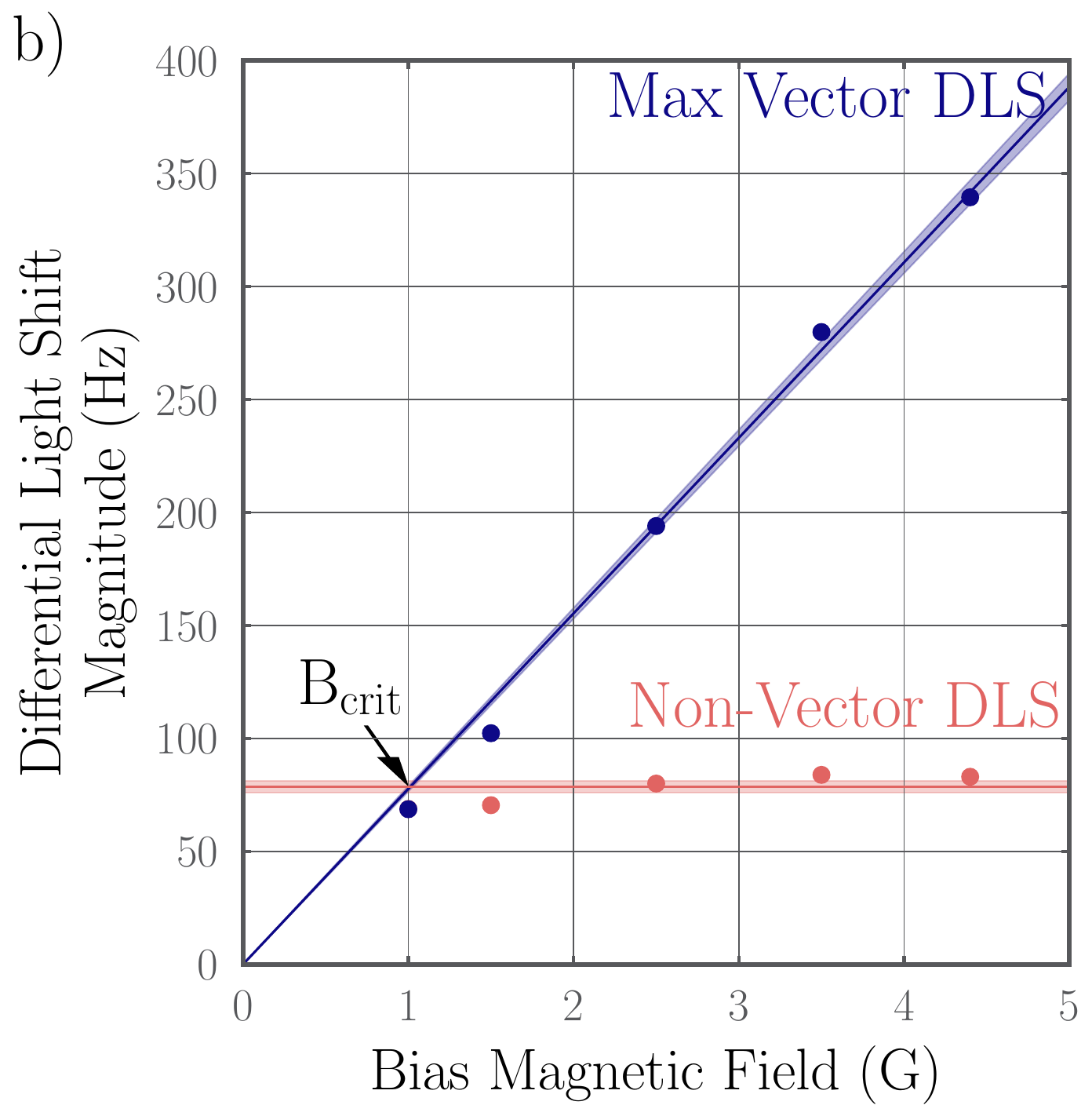}
\end{tabular}
\includegraphics[width=.9\textwidth]{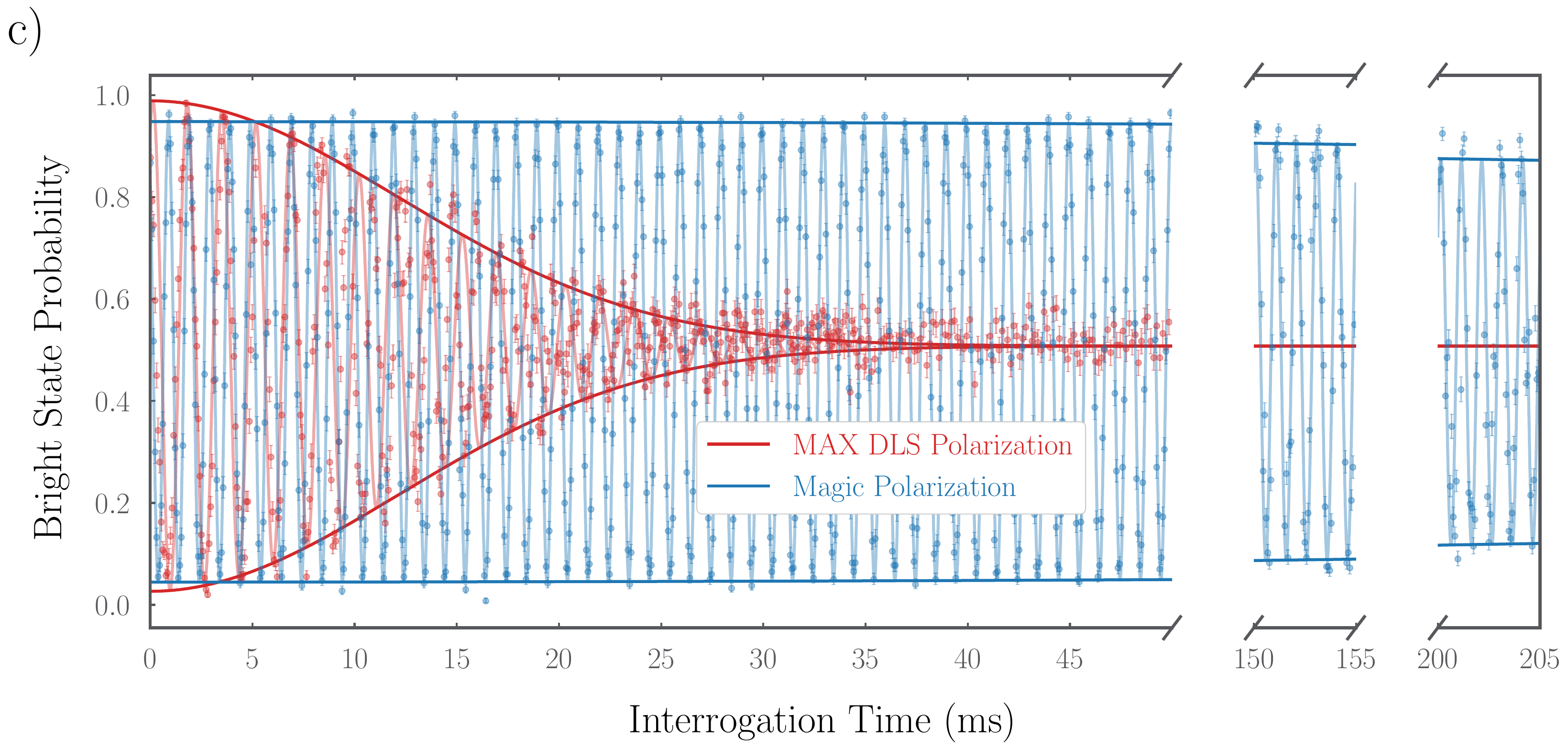}
\caption{a) A polarization-dependent DLS in the ${}^2\mathrm{S}_{1/2}$ $m_F=0$ clock qubit of \Yb{171} as a function of QWP angle at different bias magnetic field magnitudes. The $\boldsymbol{\times}$ markers denote magic polarizations for which the DLS is zero. b) Values of the the non-vector and maximum-vector DLS magnitudes with respect to bias magnetic field magnitude. The critical field is calculated from the intersection of these values as $B_{\mathrm{crit},g} =1.01 \pm 0.04 \,\mathrm{G}$. c) Detuned microwave Ramsey scans with the perturbing laser on at a maximum DLS inducing polarization vs the magic polarization shows an order of magnitude improvement to coherence time, achieving a value comparable to the control scan.
}
\label{groundmagic}
\end{figure*}

To prepare the other qubit state $^2\mathrm{F}^o_{7/2}\ \ket{3, \,0}$, we perform a second heralding sequence after preparing $^2\mathrm{F}^o_{7/2}\ \ket{4, \,0}$. Another $m$-type qubit $\pi$-pulse prepares $^2\mathrm{F}^o_{7/2}\ \ket{3, \,0}$, and a second $760\,\mathrm{nm}$ laser applied for $50\,\mathrm{ms}$ drives any population remaining in $^2\mathrm{F}^o_{7/2}\ \ket{4,\,0}$ to $^1[3/2]^o_{3/2}\; F=2$ where it decays to the ground state. Fluorescence detection in the $^2\mathrm{S}_{1/2}$ manifold then indicates failed preparation of $^2\mathrm{F}^o_{7/2}\ \ket{3, \,0}$.

Readout at the end of an experiment involves first applying a $760 \,\mathrm{nm}$ laser pulse for $50 \,\mathrm{ms}$ to drive population out of $^2\mathrm{F}^o_{7/2}\ \ket{3, \,0}$, followed by $369\,\mathrm{nm}$ fluorescence detection in the $^2\mathrm{S}_{1/2}$ manifold for $3.5\,\mathrm{ms}$. After readout, any population remaining in the $^2\mathrm{F}^o_{7/2}$ manifold is returned to the Doppler cooling cycle with both $760 \,\mathrm{nm}$ lasers applied for $50 \,\mathrm{ms}$. To repump the decay from the $^1[3/2]^o_{3/2}$ manifold to $^2\mathrm{D}_{5/2}$, $760 \,\mathrm{nm}$ lasers are always accompanied by a $976\,\mathrm{nm}$ laser.

State detection histograms for each target state prepared using this procedure are shown in Fig.\ \ref{Mqubit}(a), with high-contrast single-qubit rotations on the $m$-type qubit shown in (b).  The average $m$-type qubit state preparation and measurement (SPAM) infidelity using the two-step heralding outlined above is $2.9^{+3.0}_{-1.5}\times10^{-4} = -35\pm 4\,\mathrm{dB}$. This infidelity is limited by off-resonant driving of $^2\mathrm{F}^o_{7/2}\ \ket{4, \,0}$ to $^1[3/2]^o_{3/2}\; F=2$ during readout, resulting in a dark state reading as bright. 

\section*{\centering Magic Polarization in the \Yb{171} ground $^2\mathrm{S}_{1/2}$ qubit}

 The ground-state (``$g$-type'') clock qubit of \Yb{171} is defined in the $^2\mathrm{S}_{1/2}$ hyperfine manifold between $\ket{0,\,0}$ and $\ket{1,\,0}$, with frequency splitting $\omega_g\approx2\pi\times12.64\,\mathrm{GHz}$. Doppler cooling, optical pumping, and state readout are operated the same as in ref. \cite{YbFrequencies}. 

Differential light shifts are applied with a continuous-wave laser at $\lambda = 532 \, \mathrm{nm}$ with a $1/e^2$ intensity radius $w_0 \approx 35 \,\upmu \mathrm{m}$ at the ion. We measure differential light shifts using a modified detuned microwave Ramsey sequence in which the perturbing laser is on during the Ramsey delay time. We interleave these modified differential light shift (DLS) Ramsey scans with control Ramsey scans in which the perturbing laser is off. We fit the frequencies of both Ramsey signals to compute the differential light shift.

The perturbing laser transmits through a polarizing beam splitter for linear input polarization to a zero-order quarter wave plate (QWP). Rotation of the QWP creates polarization of sinusoidally oscillating helicity, shown in Fig.\ \ref{groundmagic}(a). The bias magnetic field points parallel to the perturbing laser $\mathbf{\hat{k}}$ such that $\pi$ polarized light is not realizable. We rotate the QWP through $180^\circ$ and measure the DLS induced by $1.0 \,\mathrm{W}$ of the perturbing laser, repeating these measurements for 5 different bias magnetic fields. We fit these data (Fig.\ \ref{groundmagic}(b)) and find $B_{\mathrm{crit},g} = 1.01 \pm 0.04 \,\mathrm{G}$ from the values of the non-vector (scalar and tensor) and maximum-vector DLS. 

To demonstrate preservation of qubit coherence, we compare the coherence times of three different detuned microwave Ramsey scans, shown in Fig.\ \ref{groundmagic}(c). One scan is performed with the perturbing laser on at a maximum DLS-inducing polarization, a second with the perturbing laser on at the magic polarization, and a control scan with the perturbing laser off. All three scans are performed with the same bias magnetic field of $5\,\mathrm{G}$. We observe the coherence time with the maximum DLS-inducing polarization to be $17.2 \pm 0.2 \,\mathrm{ms}$, while the magic polarization and control scan are comparable to one another and an order of magnitude longer. Auxiliary experiments probing the residual light shift between the control and the magic polarization scans found a residual light shift of 0.21(4)~Hz, representing a suppression of the DLS by a factor of $2000\pm400$. From this suppression of the DLS and a decoherence model assuming Gaussian amplitude fluctuations of the perturbing laser, we may infer an extension in the coherence time by the same amount. 

\section*{\centering Magic Polarization in the \Yb{171} metastable $^2\mathrm{F}^o_{7/2}$ qubit}

\begin{figure}
\includegraphics[width=.45\textwidth]{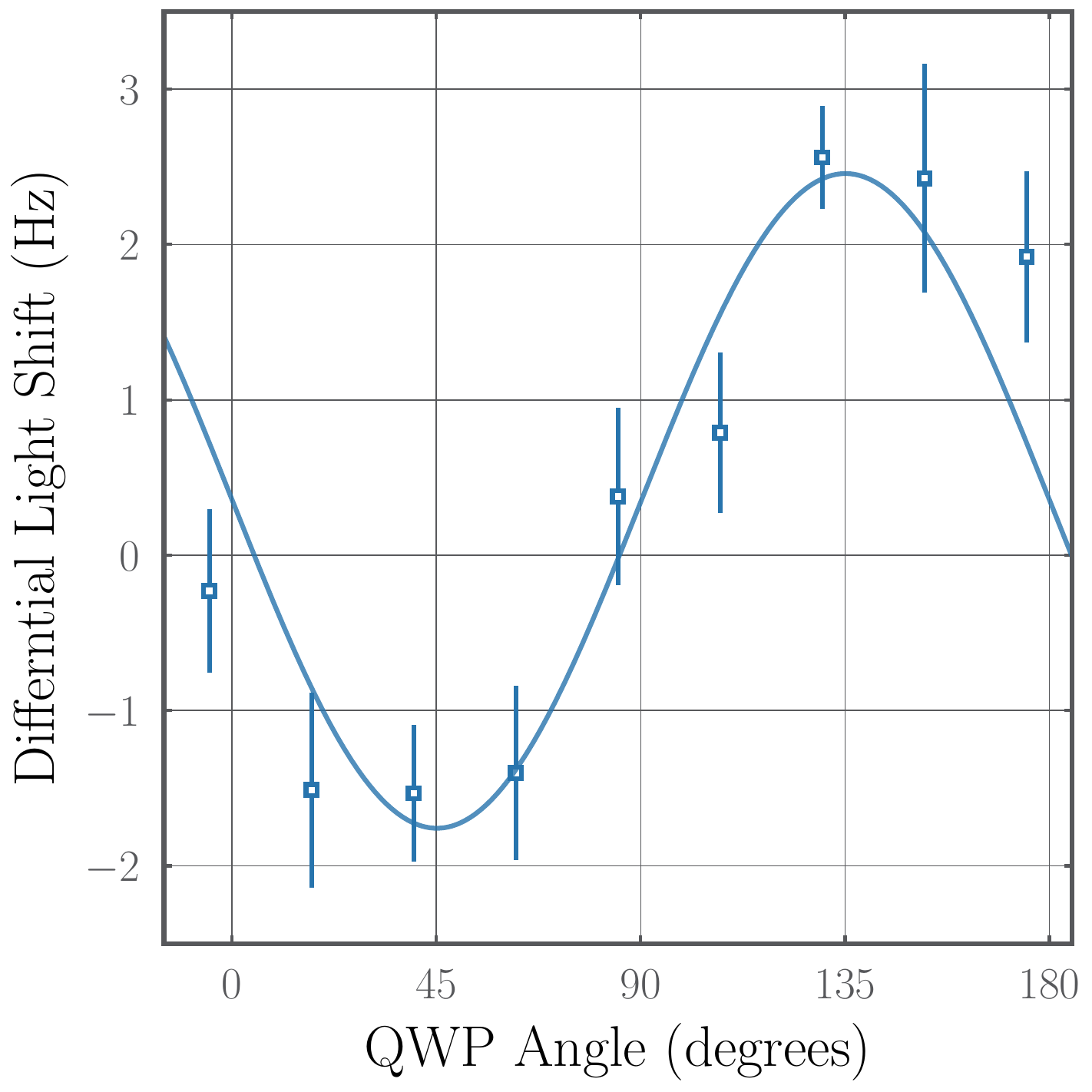}
\caption{
Polarization-dependent DLS in the ${}^2\mathrm{F}^o_{7/2}$ $m_F=0$ clock qubit as a QWP is rotated through $180^{\circ}$ with a $6.75\,\mathrm{G}$ bias magnetic field.}
\label{mQWP}
\end{figure}

We repeat a similar QWP angle experiment as in the ground state to measure the DLS in the metastable clock qubit (Fig.\ \ref{mQWP}). Due to the small magnitude of the $m$-type DLS, high measurement sensitivity is required to observe the vector light shift. We therefore increase the perturbing laser power to approximately $1.85 \,\mathrm{W}$ and increase the magnetic field magnitude to $6.75 \,\mathrm{G}$. Performing the experiment, we observe a small polarization-dependent DLS with a zero crossing ``magic'' polarization, shown in Figure \ref{mQWP}. We estimate $B_{\mathrm{crit},m} = 1.1 \pm 0.7 \,\mathrm{G}$ from the values of the non-vector and maximum vector DLS.

\section*{\centering Conclusion}

We suppress differential light shifts via magic polarization in both the ${}^2\mathrm{S}_{1/2}$ $g$- and ${}^2\mathrm{F}^o_{7/2}$ $m$- type qubits of $^{171}$Yb$^+$, and recover a coherence time in the $g$-type qubit in the presence of a perturbing laser that is comparable to the coherence time without the perturbing laser. The critical magnetic field is measured as the minimum field for which the DLS may be eliminated through magic conditions with polarization orthogonal to the quantization axis. The observed $g$-type qubit critical field $B_{\mathrm{crit},g} = 1.01 \pm 0.04\,\mathrm{G}$ is consistent with the theoretical prediction $B_{\mathrm{crit},g}=0.98\,\mathrm{G}$. For the $m$-type qubit we measure a comparable magnitude of $B_{\mathrm{crit},m} = 1.1 \pm 0.7 \,\mathrm{G}$.

Using Eq.~\eqref{B_crit_nuclear_scaling} and our measurement in $^{171}\mathrm{Yb}^+$, we estimate the critical magnetic field at $532\,\mathrm{nm}$ to be $B_\mathrm{crit}^{(173)}=0.66(3)\,\mathrm{G}$ for the ground state qubit of $^{173}\mathrm{Yb}^+$. Applying the same scaling to the $^{133}\mathrm{Ba}^+$ critical field of $1.17\,\mathrm{G}$ measured in \cite{BariumMagic}, we infer $B_\mathrm{crit}^{(135)}=0.61\,\mathrm{G}$ and $B_\mathrm{crit}^{(137)}=0.76\,\mathrm{G}$ for the ground states of $^{135}\mathrm{Ba}^+$ and $^{137}\mathrm{Ba}^+$ at $532\,\mathrm{nm}$, respectively.

The critical fields observed and inferred in this work for all isotopes of $\mathrm{Ba}^+$ and $\mathrm{Yb}^+$ are well below the field employed for typical ion trap operation, indicating that this procedure may be adopted by other researchers with minimal modifications. For all ions considered except ${}^9\mathrm{Be}^+$, the calculated magnetic field required to observe the magic polarization is below $3 \,\mathrm{G}$ even for $\lambda = 1.5\, \upmu \mathrm{m}$, and is below $10\,\mathrm{G}$ for ${}^9\mathrm{Be}^+$ for $\lambda < 800\,\mathrm{nm}$. 

\section*{Acknowledgments}
This work was supported in part by U.S.\ National Science Foundation Grants
PHY-2513134, PHY-2207985, and OMA-2016245; U.S.\ ARO grant W911NF-24-S-0004, and U.S. AFOSR grant FA9550-20-1-0323.

\nocite{roy_accurate_2017,YbFrequencies,ThomasThesis}
\bibliography{MagicRefs,library-apd,YbPlus}

\end{document}